\documentstyle[12pt,epsf]{article}
\def\la{\mathrel{\mathpalette\fun <}}
\def\ga{\mathrel{\mathpalette\fun >}}
\def\fun#1#2{\lower3.6pt\vbox{\baselineskip0pt\lineskip.9pt
\ialign{$\mathsurround=0pt#1\hfil##\hfil$\crcr#2\crcr\sim\crcr}}}
\title{Neutrino Magnetic Moment Behaviour in a Renormalizable
Model}
\author{J.M.Fr\`ere \thanks{postal address: Physique Theorique CP 225;
U.L.B. Boulevard du Triomphe; B-1050 Bruxelles, Belgium; email:
frere@ulb.ac.be}\\
Universit\'{e} Libre de Bruxelles \\
\\
R.B.Nevzorov, V.A.Novikov, M.I.Vysotsky\\
ITEP, Moscow 117259, Russia}
\date{}
\begin{document}
\large
\maketitle

\begin{abstract}

We have shown in \cite{1} that a neutrino magnetic
moment form factor
$\mu_{\nu} (q^2)$ could be
considerably amplified at low momentum transfer,
$q^2 \leq m_N^2$,
at the cost of introducing
an extra light neutral fermion $N$ with mass $m_N$ and
 nonzero magnetic moment.
It was assumed that the magnetic moment of $N$
would originate
in a renormalizable way at a heavy scale $M$.
While the enhancement of the neutrino magnetic
 moment was unambiguous,
we stressed that in this effective Lagrangian
 approach an uncertainty
persisted about the behaviour of $\mu_{\nu} (q^2)$
in the interval
$m_N^2 \ll q^2 \ll M^2$ . This is not unexpected
 in presence  of a nonrenormalizable
effective theory (a particle with
bare magnetic moment). We show in a
simple renormalizable
model for the magnetic moment of particle $N$ how a 2
loop calculation solves the ambiguity.
In  the domain $m_N^2 \ll q^2 \ll M^2$
we confirm the result obtained
in \cite{1} using a sharp cut-off .
It is amusing that the correct results are given,
as expected, through dimensional
regularization in the full theory, but NOT in
the effective lagrangian approach.
\end{abstract}

\newpage
\section{Introduction}

Paper \cite{1} considered the possibility of a
rapid variation of the
 neutrino magnetic
form factor at small momentum transfer.
Such a  rapid variation would allow comparatively
large magnetic
(transition) moments at negligible
momentum transfer
(absorption/emission of almost real photon
 by neutrino and neutrino
propagation in external magnetic field)
 despite severe bounds
from $\nu e$-scattering experiments and
the lack of pair production at
$e^+ e^- $ colliders.

A neutrino magnetic form factor
appears at the 1-loop level and varies
with momentum transfer with a typical
scale set by the virtual particles masses.
In scattering experiments momentum
transfer varies from several MeV
(for $\bar{\nu}_e$ from reactors) to
100 MeV ($\nu_{\tau}$ scattering),
and a low-energy enhancement can only
be achieved if  very
light particles run in the loop
and couple to the photon. The only existing
charged particle with a mass in the
right range is the electron, and it indeed
couples with $\nu_e$. However, the
real momentum scale is set
by the $W$-boson running in the graph,
and its huge mass determines the
value of $q^2$ at which the neutrino magnetic
form factor starts to diminish.
The way out of this was suggested in ref
\cite{1} and consists in
using a virtual neutral light spinor particle
with nonzero magnetic moment and
a light scalar instead of
charged particles in the loop.
Such particles could
have escaped detection, and the origin
of the magnetic moment of the
new fermion (a non-renormalizable interaction)
 could be found at a much
higher scale, involving particles
of mass $M \gg M_W$.

The result of the calculation of the
 neutrino magnetic form factor was
given  in \cite{1} in the following form:
\begin{equation}
\mu_{\nu}= \frac{f^2 \mu_N}{16\pi^2}
I(m_N, m_{\varphi}, m_{\nu},
q^2) \;\; ,
\label{1}
\end{equation}
where $\varphi$ is the light scalar
particle while $f$ is Yukawa
coupling constant. The interaction $f
\bar{\nu} N \varphi$ induces
$\mu_{\nu}$ at one loop
(in this paper we
consider both $\nu$ and $N$ as Dirac fermions)
(this coupling breaks the elecroweak SU(2)
symmetry, but this
can indeed occur after
the usual standard model symmetry breaking).
 In general $I \sim 1$, but in the special
case where $\nu$ and $N$
are nearly degenerate while the scalar
mass is negligeable,
($m_{\nu} \approx m_N
=m$, $m_{\varphi} \ll m_{\nu} -m_N$)
a logarithmic enhancement takes place
at small
$q^2$  \cite{1}:
\begin{equation}
I = 4\ln \frac{m}{2(m_N -m_{\nu})} -\frac{7}{2}
\;\; , \;\; q^2 \la m^2
\label{2}
\end{equation}
\begin{equation}
~~~~I = -1/2 ~~~~~~~~~~~~~~~~~~~~~~ ,
\;\; q^2 \gg m^2 \;\; .
\label{3}
\end{equation}

When deriving the expressions
(\ref{2}) and (\ref{3}) we dealt both with
ultraviolet-convergent and
-divergent integrals
over the virtual particles 4-momentum $k_{\mu}$.
The ultraviolet
divergent integral has the following form:
\begin{equation}
J_{\mu} = \int \frac{d^4 k}{(k^2 +a^2)^3} \hat{k}
\sigma_{\mu\nu}\hat{k} q_{\nu} \;\; ,
\;\; \sigma_{\mu\nu}
=\frac{1}{2}(\gamma_{\mu}\gamma_{\nu} -
\gamma_{\nu}\gamma_{\mu}) \;\;
.
\label{4}
\end{equation}

We calculated this integral in 4 space-time
 dimensions with a sharp
symmetrical cut-off
$\Lambda$ and got zero when averaging over the
directions of 4-momentum
$k_{\mu}$:
\begin{equation}
\gamma_{\alpha} \frac{1}{2}(\gamma_{\mu}\gamma_{\nu}
-\gamma_{\nu}\gamma_{\mu})\gamma_{\alpha} =
2g_{\mu\nu} -2g_{\nu\mu}
=0
\label{5}
\end{equation}
We also remarked that using instead
 dimensional regularization adds
a  non-trivial  constant due to a $d-4$ factor
 from the $\gamma$-matrices
algebra
($\gamma_{\alpha}\frac{1}{2}(\gamma_{\mu}\gamma_{\nu}-
\gamma_{\nu}\gamma_{\mu})\gamma_{\alpha} = (d-4)
\frac{1}{2}(\gamma_{\mu}\gamma_{\nu} -
\gamma_{\nu}\gamma_{\mu})$)
multiplies the pole in $1/(d-4)$ from
the integration over $|k|$.
So, in this way of
performing calculations the function $I$
shifts by a constant.
This is of course neither unexpected nor
discouraging, since we are
dealing with the effective interaction
of a particle
$N$ with bare magnetic moment, which
is nonrenormalizable.
In  \cite{1} it was naturally assumed
that the $N$ magnetic moment
was due in turn to a
loop with virtual heavy charged
particles of mass $M$, $M\ga 100$
GeV. At scale $M$ we have a perfectly
renormalizable theory while
for $q\ll M$ we
had to  deal with a nonrenormalizable
effective theory.
In such an effective theory  the
logarithmic term in (\ref{2}) which
originates from small momentum of
virtual particles is calculated
unambiguously while the precise value
of the constant may
depend on how the theory looks at the scale $M$.

Quite intriguing is the fact that the
dimensional regularization
of the effective one loop calculation
discussed above produces exaclty an
additional factor $\delta I =+1/2$,
which would result in a precocious
decrease of the magnetic moment,
long before the $M$ scale is reached:
$I \sim m^2/q^2$ for $q^2 > m^2$ !

We thus have three possibilities:

a) The constant term in $I$ depends on
the form of the theory at scale
$M$ at which magnetic moment of
particle $N$ is determined;

b) The constant term is universal
and equals 0 (as dimensional
regularization suggests),
leading to precocious power suppression
of the magnetic form factor

c) The constant term is universal
and equals $-1/2$ as suggested by
our naive sharp symmetric cut-off,
that is the superficially ultraviolet
divergent contribution from the loop vanishes identically.

To determine which of these possibilities is realized,
we study in the present note the
simplest renormalizable model for inducing the
magnetic moment  of $N$:
we assume a simple Yukawa coupling of $N$
with charged spinor and scalar
fields of equal masses $M$.

In the next Section we will calculate the
magnetic moment of $N$ in this
model. This will also bring us to discuss
the charge form factor of $N$.
The 2-loops diagrams which produce the
ordinary neutrino magnetic
formfactor will be calculated in the third Section.
Finally a comparison with the result obtained in
Sect. 2 will allow us to present expressions for the
neutrino magnetic
formfactor in a form, given by eq. (\ref{1}).
We will see that option
(c) realizes; which confirms the exact
expression given in \cite{1}.

At the end of this introduction let us make
the following technical
remark: as we are interested here only in
the neutrino magnetic form factor
behavior in the domain $m^2 \ll q^2 \ll M^2$ we will
neglect the masses of $\nu$ and of $N$ in what follows.

\section{Electromagnetic form factors of the particle $N$.}

The coupling of $N\bar{N}$ pair with a photon
is generated by two one-loop
diagrams (see Fig. 1). The corresponding
amplitude is (in order to
simplify our formulas we take the masses
$M$ of the heavy virtual spinor and scalar
particles to be equal):
\begin{eqnarray}
M_{\mu} &=&\int \frac{d^4 p(i)^6\sqrt{4\pi\alpha}}{(2\pi)^4 (p^2
-M^2)} \bar{N}\frac{1}{\hat{p}+\hat{k}_2 -M} \gamma_{\mu}
\frac{1}{\hat{p} +\hat{k}_1 -M} N+ \nonumber \\
& + &\int \frac{d^4 p (i)^6
\sqrt{4\pi\alpha}}{(2\pi)^4[(p-k_1)^2 -M^2][(p-k_2)^2 -M^2]}
\times
\nonumber \\
&\times & \bar{N}\frac{1}{\hat{p} -M} N(2p -k_1 -k_2)_{\mu} \;\; ,
\label{6}
\end{eqnarray}
where for simplicity we also put the
Yukawa coupling equal to unity.
In what
follows we will neglect mass of the
external particle $N$. Performing
integration we obtain the charge and
magnetic formfactors of $N$.
Here we meet with the first surprise
-- due to the loop correction $N$
seems to get finite non-zero charge.
Such a behavior is obviously
forbidden by
charge conservation. However,
multiplying $M_{\mu}$ by $(k_1
-k_2)_{\mu}$ to check electromagnetic
current conservation we
encounter differences of linear
divergent integrals. So in spite of
finiteness of the result for the
amplitude $M_{\mu}$, to preserve gauge
invariance we  need a proper
regularization scheme. Using dimensional
 regularization we automatically obtain
zero charge for $N$. After simple
transformations using Feynman
parametrization for the propagators we get:
\begin{eqnarray}
M_{\mu} &=& -i \int \frac{2ydydxd^d p\sqrt{4\pi\alpha}}{(2\pi)^4 [p^2
+2x(1-x)y^2(k_1 k_2)+M^2]^3} \times \nonumber \\
&\times& \bar{N}[\frac{4-d}{d}\gamma_{\mu}p^2 -M^2 \gamma_{\mu}
-2x(1-x)y^2 \times \nonumber \\
& \times & (k_1 k_2)\gamma_{\mu} +4x(1-x)y^2(k_1 k_2)\gamma_{\mu}
+(k_1 +k_2)_{\mu} M]N \;\; .
\label{7}
\end{eqnarray}

The sum of the first three terms in brackets is zero
 the fourth term
generates a charge form factor while
the last term is the magnetic
form factor we want to study
(for massless fermions $(k_1 +k_2)_{\mu}\bar{N} N =
\bar{N}\sigma_{\mu\nu}q_{\nu} N$).
For small momentum transfer
($q^2 \equiv (k_1 -k_2)^2 \ll M^2$) we get:
\begin{equation}
M_{\mu} =\frac{i\sqrt{4\pi\alpha}}{192\pi^2} \frac{q^2}{M^2}
\bar{N} \gamma_{\mu} N -\frac{i\sqrt{4\pi\alpha}}{32\pi^2 M}
\bar{N} N (k_1 +k_2)_{\mu} \;\; ,
\label{8}
\end{equation}
here the first term describes the charge radius
of particle $N$, while
the second describes its magnetic moment.

To end this Section let us note that for
large momentum transfer $q^2
\gg M^2$ the magnetic form factor falls
down as expected like$\sim M^2/q^2
\ln^2(q^2/M^2)$, while the charge form factor
tends to a (non-zero)
 constant.
(It is interesting to consider the  charge form
factor in the light
of dispersion relations. As the imaginary part of
the form factor falls down,
an unsubtracted dispersion relation can be written.
It
produces a form factor which falls down for
$q^2 \gg M^2$, but contains
nonzero charge for $N$. Subtracting this
"charge" we get constant
behavior for $q^2 \gg M^2$.)

\section{Neutrino magnetic form factor for $q^2 \gg m^2$.}

The two Feynman diagrams shown in Fig. 2
are in turn responsible for the
ordinary neutrino
magnetic form factor. Since we neglect
both $\nu$ and $N$ masses, helicity
flip (necessary for a magnetic form factor)
can only occur in the inner
fermion line (with mass $M$). As the first
part of calculation we perform
the integral over momentum $p$. Taking into
account only the
 terms proportional to
$M$ (they are the only ones contributing to
the  magnetic moment)
 we obtain:
\begin{equation}
A_{\mu}=i \int \frac{2ydydx}{32\pi^2}\frac{M[\hat{k}_2\gamma_{\mu}
+\gamma_{\mu}\hat{k}_1 -(k_1 +k_2)_{\mu}] \sqrt{4\pi\alpha}}
{[M^2 -y(1-y)(k-xk_2 -(1-x)k_1)^2 +2yx(1-x)k_1 k_2]} \;\; ,
\label{9}
\end{equation}
where $0 <y,x <1$. When expression
(\ref{9}) is inserted into  the
external loop we get as expected
from renormalizability an
extra suppression of the integrand over $k_{\mu}$ in
the domain $k^2 > M^2$. Taking into
account the propagators of fields
$\varphi$ and $N$ we observe that
the integral over $k_{\mu}$ is now u.v.
convergent  by power counting.
In this way, our microscopic
model of the $N$ particle magnetic moment
regularizes the expression for
the light neutrino magnetic form factor.
This leads as announced  to the convergence of the
analog of integral (\ref{4}) in  the
2-loop approach;
 so even making  {\it the full integration}
in space-time dimensions
$d \neq 4$ we get zero for  the
integral proportional to $\hat{k}\sigma_{\mu\nu}\hat{k}$.
In this way
the result obtained in \cite{1} for the ordinary neutrino
magnetic moment is
justified; our option c) realizes.

Armed with the qualitative arguments given
above let us proceed with
the calculation of the second loop.
Making use of Feynman parametrization
for propagators of particles $N$ and
$\varphi$ (we use parameters $u$
and $v$, $0 < u$, $v < 1$) after
simple transformations we get:
\begin{equation}
T_{\mu} = f^2 \int \frac{2ydydx \sqrt{4\pi\alpha}}{32\pi^2}
\frac{d^4 k2v dvdu}{(2\pi)^4[k^2 -2k_1 k_2 uv^2(1-u)]^3}
\times
\label{10}
\end{equation}
$$
\times \frac{M\bar{\nu}_2 [-(k_1 +k_2)_{\mu}2k_1 k_2 u(1-u)v^2
+\hat{k}((k_1 +k_2)_{\mu} -\hat{k}_2 \gamma_{\mu}
-\gamma_{\mu}\hat{k}_1)\hat{k}]\nu_1} {[M^2 -y(1-y)(k-xk_2
-(1-x)k_1 + uvk_2 +(1-u)v k_1)^2 +2xy(1-x)k_1 k_2]} .
$$

The second term in brackets in the
nominator produces corrections to
the
neutrino magnetic moment of order $\frac{1}{M}\times
\frac{q^2}{M^2}$ and is negligible for $q^2 \ll M^2$.
To calculate
the
contribution of the first term, let us begin
with comparatively small
momentum transfer, $q^2 \ll M^2$. For such momentum
transfer the
second bracket in the denominator equals $M^2$
(let us mention that
integral over $k_{\mu}$ is u.v. convergent for the part of
$T_{\mu}$ now considered) and we readily get:
\begin{equation}
T_{\mu} =\frac{i\sqrt{4\pi\alpha}f^2}{32\pi^2 M}\bar{\nu}_2 \nu_1
(k_1 +k_2)_{\mu} \frac{1}{32\pi^2} \;\; , \;\; \mbox{\rm or} \;\;
I=-\frac{1}{2} \;\; \mbox{\rm for} \;\; q^2 \ll M^2 \;\; .
\label{11}
\end{equation}

For large momentum transfer, $q^2 \gg M^2$
the second bracket in
denominator in (\ref{10}) is proportional to
$q^2$ which means that  the
neutrino magnetic moment falls down,
$I\sim M^2/q^2$ according to
general expectations for a renormalizable field theory.

\section{Conclusions.}

In this paper we demonstrated that an
effective theory with a
neutral spinor
particle $N$ with nonzero magnetic moment allows to
calculate the
induced
neutrino magnetic form factor in the domain $q^2 \ll M^2$,
where $M$
is scale at which $N$ magnetic moment is generated if loop
calculations are made with naive ultraviolet cut-off.
This is
fully consistent with a dimensional regularization treatment {\it
of the full renormalisable underlying theory},
but {\it not} of the
effective model.

All authors were supported by INTAS grant 94-2352;
RBN, VAN, MIV acknowledge support of INTAS
grants INTAS-93-3316-ext,
INTAS-RFBR 95-0567 and grant RFBR-96-02-18010 as well.

\newpage

\begin{figure}
\epsfbox{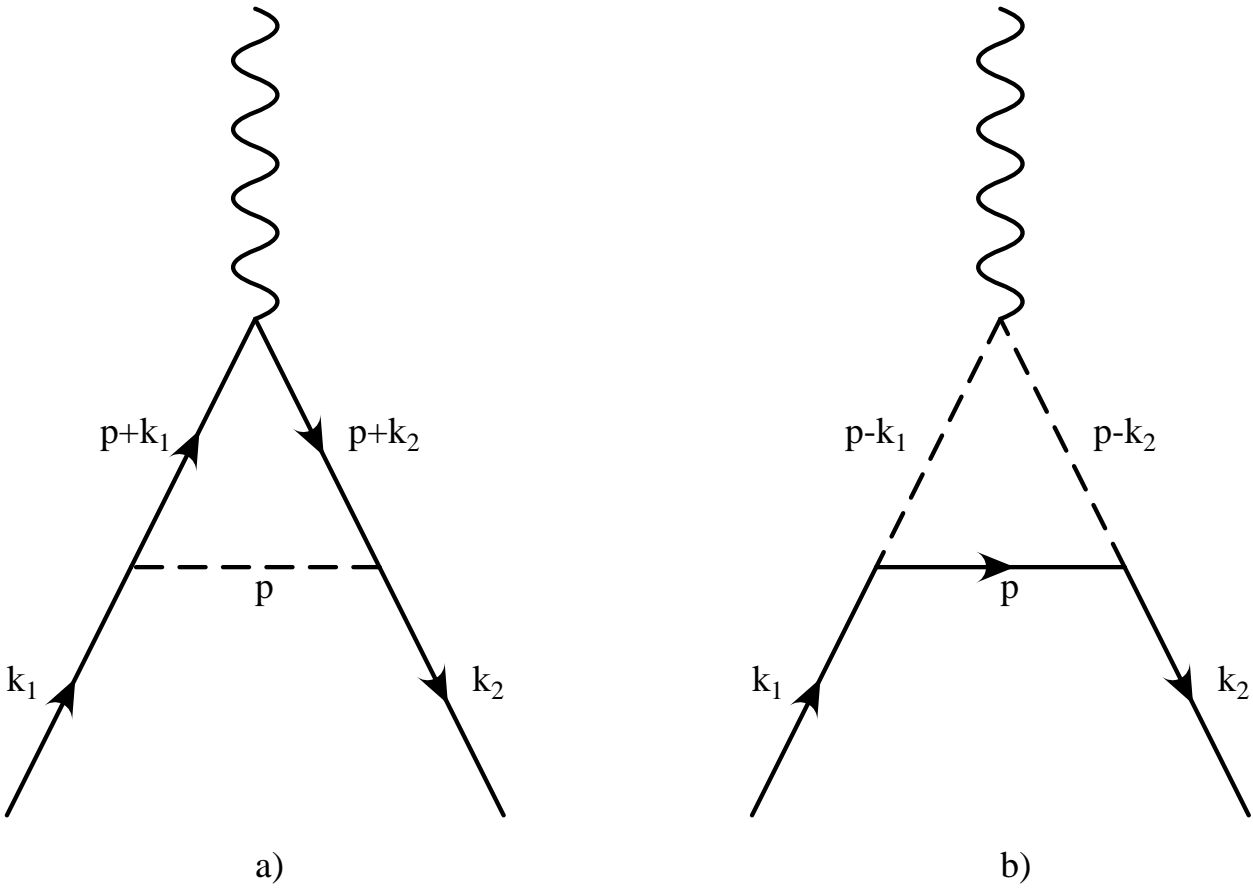}
\caption{These diagrams generate the coupling of a $N\bar{N}$ pair with a
photon.}
\end{figure}

\newpage

\begin{figure}  
\epsfbox{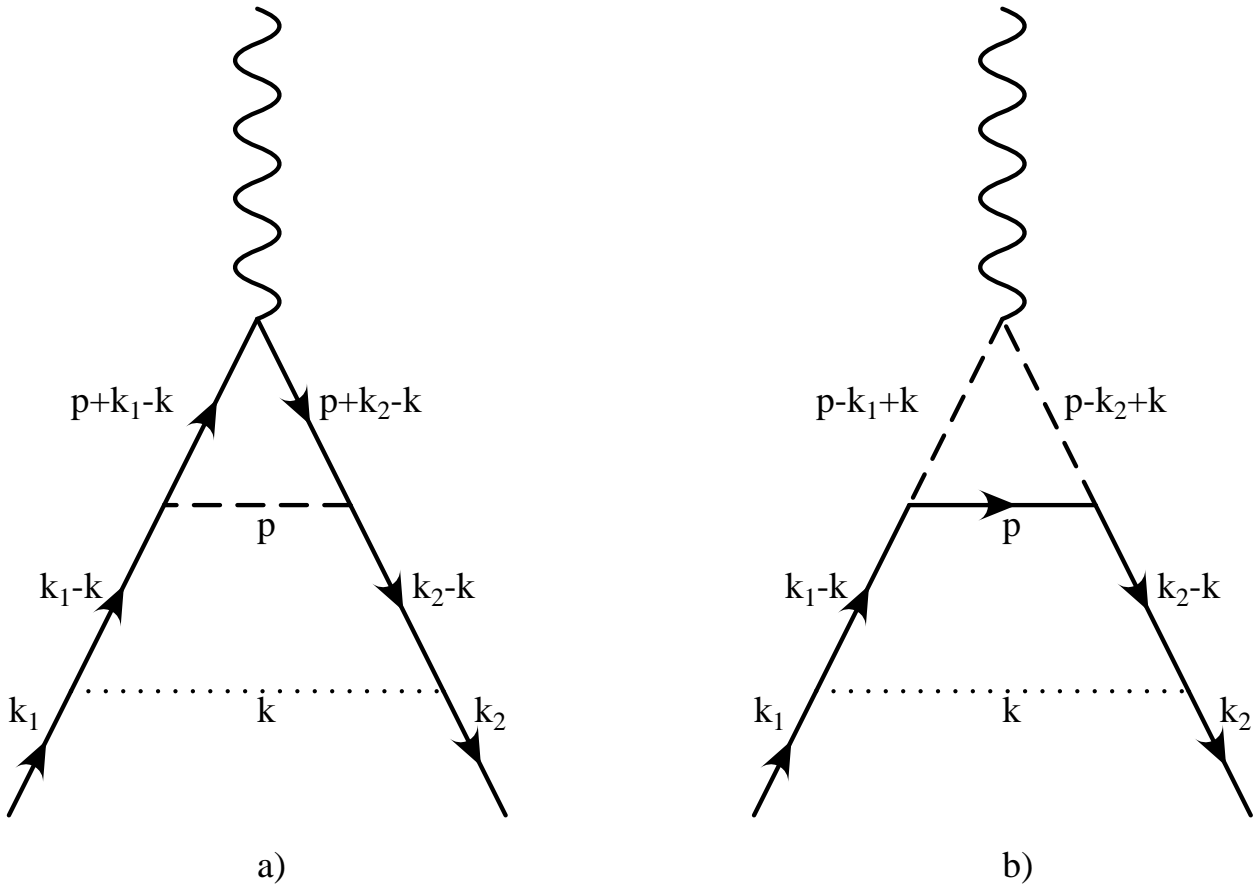}
\caption{These two--loop diagrams contribute to the neutrino magnetic form
factor.}
\end{figure}

\end{document}